\documentclass[conference]{IEEEtran}
\IEEEoverridecommandlockouts
\usepackage{cite}
\usepackage{amsmath,amssymb,amsfonts}
\usepackage{algorithmic}
\usepackage{graphicx}
\usepackage{textcomp}
\usepackage{xcolor}
\usepackage{mathtools}
\def\BibTeX{{\rm B\kern-.05em{\sc i\kern-.025em b}\kern-.08em
    T\kern-.1667em\lower.7ex\hbox{E}\kern-.125emX}}
\begin{document}

\title{Analyzing GPU Tensor Core Potential for Fast Reductions\\
\thanks{}
}

\author{
\IEEEauthorblockN{Roberto Carrasco}
\IEEEauthorblockA{\textit{Instituto de Inform\'atica} \\
\textit{Universidad Austral de Chile}\\
Valdivia, Chile \\
rcarrasco@comafor.cl}
\and
\IEEEauthorblockN{Raimundo Vega}
\IEEEauthorblockA{\textit{Instituto de Inform\'atica} \\
\textit{Universidad Austral de Chile}\\
Valdivia, Chile \\
}
\and
\IEEEauthorblockN{Crist\'obal A. Navarro}
\IEEEauthorblockA{\textit{Instituto de Inform\'atica} \\
\textit{Universidad Austral de Chile}\\
Valdivia, Chile \\
}
}
\IEEEoverridecommandlockouts
\IEEEpubid{\makebox[\columnwidth]{978-1-5386-9233-2/18/\$31.00~\copyright2018 IEEE \hfill} \hspace{\columnsep}\makebox[\columnwidth]{ }}
\maketitle

\begin{abstract}
The Nvidia GPU architecture has introduced new computing elements such as the \textit{tensor cores}, which are special processing units dedicated to perform fast matrix-multiply-accumulate (MMA) operations and accelerate \textit{Deep Learning} applications.
In this work we present the idea of using tensor cores for a different purpose such as the parallel arithmetic reduction problem, and propose a new GPU tensor-core based algorithm as well as analyze its potential performance benefits in comparison to a traditional GPU-based one. The proposed method, encodes the reduction of $n$ numbers as a set of $m\times m$ MMA tensor-core operations (for Nvidia's Volta architecture $m=16$) and takes advantage from the fact that each MMA operation takes just one GPU cycle. When analyzing the cost under a simplified GPU computing model, the result is that the new algorithm manages to reduce a problem of $n$ numbers in $T(n) = 5\log_{m^2}(n)$ steps with a speedup of $S = \frac{4}{5}\log_2(m^2)$.
\end{abstract}

\begin{IEEEkeywords}
Reduction; NVIDIA Tensor Cores; GPU Computing; matrix-multiply-accumulate;
\end{IEEEkeywords}

\section{Introduction}
In the present day, the amount of data generated by all sorts of media grows at an exponential rate, making each year more difficult to process these data sets entirely. On top of this, the rise of applications such as autonomous vehicle platforms, pattern recognition in images, disease diagnosis, weather forecasting, financial modeling, robotic and language translation, among others impose an ever greater challenge which is to perform in real-time. One important computational pattern in many applications is the arithmetic reduction which computes a single value from a set of $n$ elements. Arithmetic reductions are employed in many of the applications recently mentioned as well as in Monte Carlo methods, physical simulations such as the gravitational n-body problem and ray tracing, among many others. Therefore, it is of high interest to count with a fast reduction method and give the possibility to reach real-time performance in these applications.

\subsection{The GPU Programming Model}
Today GPU Computing \cite{navarro_hitschfeld-kahler_mateu_2014} is a useful computational tool for attempting the solution to large problems in the big data era, as it offers massively parallel computation for data-parallel problems. This means that  elements in a large data-set can actually be processed by a parallel algorithm if the problem can be defined as data-parallel. GPU Computing has been used successfully in the recent years, giving important performance speedups in physics \cite{CARTER2018148, NAVARRO201648}, all-pairs problems \cite{8392762}, medicine \cite{0031-9155-56-22-002}, image processing \cite{CERDA20188}, deep learning \cite{SCHMIDHUBER201585} and computer graphics \cite{Chaitanya2017}, among other fields.

\begin{figure}[ht]
\centerline{\includegraphics[scale=.25]{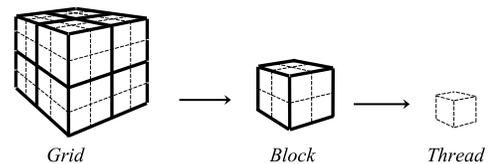}}
\caption{The three-level hierarchy of the GPU Programming.}
\label{fig_constructs}
\end{figure}

When working with GPU Computing, its programming model plays an important role in the design and development of GPU accelerated programs, and it is governed by a three-level hierarchy, which defines the parallel space of the GPU. This hierarchy corresponds to the thread, block and grid (See Figure \ref{fig_constructs}). A thread is an abstract compute element that is in charge of executing the kernel once. A block is a container of threads, and has the property that all of its threads can synchronize among themselves  and can share information via cache. Synchronization among blocks is not possible in the current programming model, unless the kernel is terminated. The grid is the last construct, and it contains all the blocks, spatially identified, that will be employed in the execution of the kernel. Lastly, the kernel is the function that will execute in GPU by all threads. With the help of programming tools such as OpenCL or CUDA, one can implement a parallel GPU algorithm that will accelerate the data-parallel computations.

\subsection{GPU Tensor Cores}
\begin{figure}[ht]
\centering
\includegraphics[scale=.34]{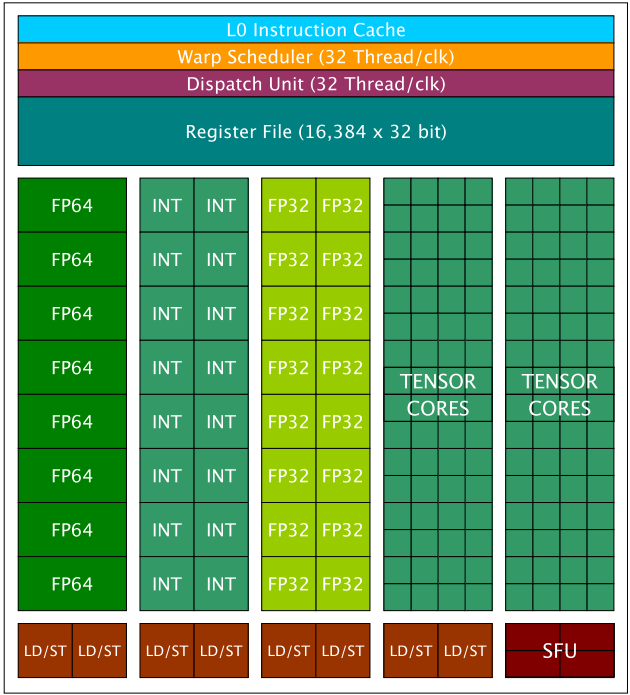}
\caption{A processing group of the Nvidia Tesla V100 which has a total of 640 tensor cores. Image inspired from the Nvidia CUDA programming guide \cite{cuda}.}
\label{fig_gpu_tensor_cores}
\end{figure}

With the latest rise of Machine Learning applications, and more specifically the fast adoption of Deep Learning in multiple fields of science and technology, CPU and GPU Companies have started including application specific integrated circuits (ASICs) to their processors to further accelerate the computational tasks involved in the phases of training and inference in Deep Learning applications. This change has led to the inclusion of \textit{Tensor Core} (TC) units in recent Nvidia GPUs, which are special purpose processing units that sit next to the GPU CUDA cores in the streaming multi-processors (SM) of the chip, as shown in Figure \ref{fig_gpu_tensor_cores}.

The aspect that make tensor cores an attractive feature is the performance it can offer in comparison to the traditional GPU cores. Today, the Nvidia Volta GPU Tesla V100, Quadro V100 and Titan V all include around 640 tensor cores, and they can offer up to 120 TFLOPS in mixed FP16-FP32 precision. In comparison, the traditional CUDA cores, which are 5120 in total for the GPUs recently mentioned, offer up to $\sim 15$ TFLOPS of performance in FP32 precision and around $\sim 7$ TFLOPS in FP64 precision. Indeed, the fast performance of tensor cores is an attractive opportunity to explore possible applications that can take advantage of this new technology. However, the use of tensor cores does not come free of restrictions and potential issues. The first restriction is that the programming of tensor cores is much more restrictive than the programming of traditional GPU cores. In fact, the only operation allowed when using tensor cores is the matrix-multiply-accumulate (MMA) in matrices\footnote{Although GPU tensor cores work at hardware level with $4\times 4$ matrices, the programming model exposes the operation in terms of $16 \times 16$ matrices.} of $4 \times 4$, \textit{i.e.}, 
\begin{equation}
    D_{4\times 4} = A_{4 \times 4} \times B_{4 \times 4} + C_{4 \times 4}
\end{equation}
which takes just one GPU cycle. Therefore, the only way to make an algorithm to take advantage of tensor core acceleration is to redesign it as a set of many $4\times 4 \times 4$ MMA operations that can run in parallel.

The second restriction of tensor cores is the numerical precision. Although the resulting $D$ matrix can be stored in FP32 precision, currently the operations $A\times B + C$ are done in FP16 precision. This mixed mode of operation may have negative effects in some applications. The reason of why tensor cores operate at FP16 is because in general Deep Learning applications do not suffer from numerical precision when doing training and inference in FP16.

Given the importance of arithmetic reductions in the era of big data and the recent advances in GPU Tensor Cores for fast MMA computations, we formulate the following research question: \textit{Can tensor core performance be exploited to accelerate arithmetic reductions and speedup  applications from science and technology?}
This work aims at answering this question by analyzing the potential performance improvement one could encounter by doing arithmetic reductions with tensor cores. For this, we propose a new parallel reduction strategy, based on a hierarchy of MMA operations. Our findings show that in theory a tensor core based reduction can be significantly faster than a traditional GPU one, as it only takes  $T(n) = 5\log_{m^2}(n)$ time steps, with $m$ being the linear 
size of the MMA matrices involved.

The rest of the manuscript presents the related work (Section \ref{sec_related_work}), an overview of the parallel reduction (Section \ref{sec_overview_parred}), the formulation of the new tensor core algorithm and its analysis (Section \ref{sec_formulation_analysis}), and a discussion of the results obtained in Section \ref{sec_discussion}.

\section{Related Work}
\label{sec_related_work}
The parallel reduction has been implemented using different frameworks. In the case of OpenMP there are high level abstract constructs that allow the programmer to express a parallel reduction via OpenMP pragma commands \cite{case_study_omp_reduce}. In the case of GPUs, the parallel reduction has been addressed by Nickolls, Buck and Garland \cite{nickolls_2008}. The authors propose a parallel sum reduction of $O(\log_2(n))$ time, where each thread loads one element of the input array, and then adds pairs of values in parallel as $x[i] = x[i] + x[i+p/2]$ where $p$ is the number of threads. The loop in this kernel implicitly builds a summation tree over the input elements, and at the end of this loop, the first data slot of each thread-block holds the reduction result of one iteration, \textit{i.e.}, $x_{B}[0] = \sum_{i=0}^{p} x_{i}$. More kernels are executed until the input problem is one value.

Harris optimized for CUDA an algorithm for parallel reduction \cite{harris_2005} tree-based as well. In his work, the author illustrates seven different optimizations that are relevant to the parallel reduction, achieving a final version that is up to $30x$ times faster than the initial GPU version presented. Harris mentions that although the time complexity of the parallel reduction is indeed $O(\log_2(n))$, the cost is not efficient if $p=n/2$ threads are used. The author shows, with the help of Brent's Theorem, that using $p = n/\log_2(n)$ threads leads to a parallel efficient cost algorithm.

Another alternative, simpler in design complexity but still competitive in performance, is to use the intrinsic Atomic Add \cite{nvidia-cublas} GPU instruction. The operation is atomic in the sense that it is guaranteed to be performed without interference from other threads, but can still run in parallel if threads are using the atomic operation on different data addresses. The atomic add reads a number at some address in global or shared memory, adds another number to it, and writes the result back to the same address, in one transaction.

In the case of distributed computing, there are two levels of parallelism that must occur for the reduction to be completed; (1) local reduction and (2) distributed reduction. For the local reduction, the process may be carried with multi-core CPU or GPU computation as recently described. For the case of distributed computation, the results of different compute nodes must be merged with message passing tools such as MPI \cite{omp_ompi}. The result is an hybrid OpenMP-MPI or GPU-MPI reduction for massive scale systems. Recent tools such as MapReduce \cite{mapreduce} also offer a higher-level of abstraction to accomplish parallel reduction in cluster environments. 

The most recent and relevant work about CUDA GPU tensor core programming is the one by Markidis \textit{et al.} \cite{tensor_cores_2018}, in which they studied current approaches to program NVIDIA Tensor Cores, as well as the performance and the precision loss due to computation in mixed precision. The authors show how NVIDIA CUDA provided three ways of programming the matrix-multiply-accumulate (MMA): CUDA Warp MMA (WMMA) API, CUTLASS, and cuBLAS GEMM. The tensor core programming is analyzed in different aspects such as programmability, performance and precision. The authors report that the maximum performance obtained was with the cuBLAS GEMM implementation, where they achieved 83 $TFLOPS$ in their test environment (approximately $74\%$ of the theoretical performance), followed by CUTLASS with 62 $Tflops/s$. The WWMA implementation did not provide any performance improvement, however the authors realized they did not use shared memory in the process, which is an important aspect in order to accomplish efficient tensor core computation. They also observed that when the size of the input matrix increases, the error may increase significantly in some cases. For this, the authors include mechanisms to increase precision such as the Kahan summation and \textit{iterative precision refinement}. 

In more general terms, the Google Tensor Processing Unit (TPU), deployed for data-centers in 2015, is another processor comparable to the tensor cores found in Nvidia GPUs, that accelerates the inference phase of neural networks  \cite{tpu_google_2017}. Norman \textit{et al.} compared the TPU to a server-class Intel Haswell CPU and an Nvidia K80 GPU. They used workloads written with the TensorFlow framework. The results showed that the TPU was on average about 15$\times$ - 30$\times$ faster than its contemporary GPU or CPU, and about 30$\times$ - 80$\times$ higher in TOPS/Watt (Tensor Operations per Second per Watt).

\section{Overview of the Classic Parallel Reduction}
\label{sec_overview_parred}
Given an array of $n$ elements,  $X = \{x_1, x_2, ..., x_n\}$,   
the arithmetic reduction $R(X)$ is defined as
\begin{equation}
    R(X) = \sum_{i=1}^n x_i
\end{equation}
On a serial processor, one would write a simple loop with a single accumulator variable to construct the sum of all elements. Such algorithm is the best sequential one (as in principle no element can be left out) and it costs $\Theta(n)$.

A parallel reduction algorithm cannot solve the problem in one time step, 
as there are dependencies and race conditions to be fulfilled within the reduction process. Nevertheless, it is known that a parallel reduction can sum pairs of values in parallel at each time step, leading to a parallel cost of $T_{p=n/2}(n) = O(\log_2(n))$ using $p=n/2$ processors. At each time step $k$, we have a problem of 
size $\frac{n}{2^{k-1}}$ and $\frac{n}{2^k}$ threads prepared to do work. Each parallel thread sums a pair of values, \textit{i.e}, the $i$-th thread sums the elements $x_i$ and $x_{i + n/{2^k}}$ and stores the partial sum in $X[i]$. Such parallel step costs $O(1)$ time and cuts the problem in half. Consecutive applications of this process lead to the recurrence
\begin{align}
T(n) &= O(1) + T(n/2)
\end{align}
subject to $T(2) = O(1)$, 
which by the master theorem it can be shown that it is $T_{p=n/2}(n) = O(\log_2(n))$. 

When considering the parallel cost, which is defined as $C_p = T_p(n) \cdot p$, with $p$ the number of processors employed, one can realize that using $n/2$ processors leads to a parallel cost of
\begin{equation}
C_p(n) = \log_2(n) \frac{n}{2} = O(n \log_2(n))
\end{equation}
which makes the algorithm cost inefficient as it is greater than the $O(n)$ cost of the sequential algorithm. To improve on this cost, one can use Brent's Theorem \cite{Brent_1974} that states the following inequality
\begin{equation}
T_p(n) \le \frac{T_1(n)}{p} + T_{\infty}(n)
\end{equation}
where in the case of the arithmetic reduction we have that $T_1(n) = O(n)$ and $T_{\infty}(n) = \log_2(n)$. With this, one can choose the number of processors as $p = \frac{n}{\log_2(n)}$ without sacrificing parallel time, \textit{i.e}, 
\begin{equation}
    T_p(n) \le 2\log_2(n) = O(\log_2(n))
\end{equation}
With this change, the cost of the parallel algorithm is now efficient, \textit{i.e},
\begin{equation}
    C_p(n) = \log_2(n) \frac{n}{\log_2(n)} = O(n)
\end{equation}

It is important to consider that there is an asymptotic lower bound of $\Omega(\log_2(n))$ for the parallel reduction. Nonetheless, one can 
still improve in the constants involved as in the base of the logarithm, which can make an important difference in experimental performance. Such is the case of reducing with tensor cores, where a large number of arithmetic operations can be encoded into MMA operations and executed in just one GPU cycle, which is equivalent to one time unit.

\section{A new Reduction Algorithm based on MMA}
\label{sec_formulation_analysis}
In this section we present the new algorithm for parallel arithmetic 
reductions using tensor cores operations and analyze its parallel time in both asymptotic and non-asymptotic forms. 

\subsection{Formulation}
The tensor core programming model exposes a single operation to the programmer, the matrix-multiply-accumulate (MMA). That is, given three matrices $A, B, C$, the MMA operation computes
\begin{equation}
    D = A \times B + C
\end{equation}
In one GPU cycle. The tensor core computing model allows many MMA operations to occur simultaneously in parallel. It is interesting to note that in the programming model the tensor core MMA operation is exposed in terms of $16 \times 16$ matrices to the programmer, even when the actual operation at hardware level is carried in terms of $4 \times 4$ matrices. The process of splitting the $16 \times 16$ workload into smaller $4 \times 4$ works is done automatically by the GPU scheduler, but splitting a large problem of size $n$ into several $16 \times 16$ matrices is not automatic and the partition must be designed manually. This last aspect is the one important for the research, as it is related to the research question of wether a reduction problem of size $n$ can be encoded into multiple MMA operations. The presentation of the new reduction algorithm will proceed in terms of $m \times m$ MMA matrices, as it favors the analysis in the next sub-section. 

The main intuition behind a tensor core MMA based reduction 
is to produce many partial summations of groups of $m^2$ numbers, in parallel. To achieve this, we employ two MMA operations. For the first MMA operation, $m^2$ elements of the input array $X[\ ]$ are inserted in $A_{m\times m}$ in parallel such that $A_{m,m}$ is actually the $m^2$-th element of the group. Then we set $B_{m \times m}$ as an all-ones matrix, also in parallel, and $C$ is a zero-matrix, also set in parallel. 
When the MMA operation is executed on $A, B$ and $C$, the result is a matrix, namely $D_{m \times m} = A_{m \times m} \times B_{m \times m} + C_{m\times m}$, of the form 
\begin{align} \label{eq_MMA1}
D &= 
\begin{bmatrix}
    x_{11} & \dots  & x_{1m} \\
    \vdots & \ddots & \vdots \\
    x_{m1} & \dots  & x_{mm}
\end{bmatrix}
\times
\begin{bmatrix}
  \mbox{\huge1}\\
\end{bmatrix}_{m \times m}
+
\begin{bmatrix}
  \mbox{\huge0}\\
\end{bmatrix}_{m \times m}
\\
&=
\begin{bmatrix}
    \sum_{i=1}^{m} x_{1i}  & \dots &  \sum_{i=1}^{m} x_{1i}  \\
    \vdots & \ddots & \vdots \\
    \sum_{i=1}^{m} x_{mi}  & \dots &  \sum_{i=1}^{m} x_{mi}  
\end{bmatrix}
\end{align}
where each column $_{k,j}$ holds the whole set of partial summations from the group of $m^2$ elements.

The second MMA operation is in charge of reducing the partial summations found in the columns of $D$, now into a single value. It is relevant to notice that dealing with one column of $D$ is sufficient has it holds all the partial summations. The other columns of $D$ hold copies of the result. At the same time, given the rigidness of the MMA operation, it takes less time to process the whole matrix within the MMA operation instead of filtering or doing extra considerations to just process a single column. As long as the result of one column is not compromised, doing a full MMA operation is still the most convenient approach in this contexts as we preserve the one GPU cycle cost per MMA. 

The second MMA operation proceeds by changing the order of the multiplying matrices, and re-uses the output matrix $D$ in the position of $B$, while using $B$ in the position of $A$. With these changes, the second MMA operation becomes the following expression 
\begin{align} \label{eq_MMA2}
D' &= 
\begin{bmatrix}
  \mbox{\huge1}\\
\end{bmatrix}
\times
\begin{bmatrix}
    \sum_{i=1}^{m} x_{1i}  & \dots &  \sum_{i=1}^{m} x_{1i}  \\
    \vdots & \ddots & \vdots \\
    \sum_{i=1}^{m} x_{mi}  & \dots &  \sum_{i=1}^{m} x_{mi}  
\end{bmatrix}
+
\begin{bmatrix}
  \mbox{\huge0}\\
\end{bmatrix}
\\
&=
    \begin{bmatrix}
    \sum_{i=1}^{m}\sum_{j=1}^{m} x_{ij}   &  \dots  & \sum_{i=1}^{m}\sum_{j=1}^{m} x_{ij}  \\
    \vdots & \ddots & \vdots \\    
    \sum_{i=1}^{m}\sum_{j=1}^{m} x_{ij}   &  \dots  & \sum_{i=1}^{m}\sum_{j=1}^{m} x_{ij}  \\
\end{bmatrix}
\end{align}
The resulting matrix $D'$ contains the reduction of the $m^2$ numbers, replicated in all of its elements. Once the second MMA operation finishes, the thread in charge proceeds to write the reduction result, \textit{e.g}, $D'_{1,1}$, into global memory 
as part of the new array of partial sums.

The global idea of the algorithm is to subdivide the domain of $n$ numbers into $\frac{n}{m^2}$ blocks of $m^2$ elements and execute the 2-step MMA reduction proposed for each group in parallel. This reduces the problem size by a factor of ${m^2}$. In the next iteration, the idea is to take the smaller version of the problem, of size $n' = \frac{n}{m^2}$ and reduce it again with the 2-step MMA operations for all the possible groups of $m^2$ elements. This process is carried iteratively until the reduction set fits in just one group of $m^2$ elements, for which a final tensor core reduction returns the result of the whole reduction problem of size $n$.

This new tensor core MMA based reduction, namely $R_{tc}(X)$, with $X = \{x_1, \dots, x_n\}$, can be described by the following recurrence
\begin{align}
R_{tc}(X) &= R_{tc}(M(x_1..x_{m^2}),\dots, M(x_{(k-1)m^2+1}..x_{km^2}))
\end{align}
where $M(...)$ is the tensor core based MMA reduction and the initial condition 
is defined as
\begin{equation}
    R_{tc}(x_1..x_{m^2}) = M(x_1..x_{m^2})
\end{equation}
In the following subsection we perform a fine analysis (considering constants) of the performance of the tensor core based reduction algorithm.

\subsection{Analysis of Performance}
In order to analyze the cost of the tensor core based reduction algorithm, we utilize a simplified GPU Computing model similar to the PRAM but with extra restrictions related to the GPU architecture. In this simplified model, the costs for the different types of parallel operations are:
\begin{itemize}
    \item Coalesced read/write operations cost $1$ unit of time.
    \item non-Coalesced read/write operations cost $w$ units of time. 
    \item Parallel tensor core MMA operations cost $1$ GPU cycle.
    \item Simultaneous r/w into tensor core matrices costs $1$ unit of time.
\end{itemize}

In the case of a GPU reduction, it is possible to produce coalesced memory acceses, by not using a stride in the access pattern, and instead make threads access the array $X$ in blocks without gaps among them. Once threads read their corresponding data, which takes one unit of time, they can load this data into the tensor core matrices, which reside in cache memory near the tensor cores. This loading of information takes another unit of time. Then, the algorithm proceeds and executes the 2-step tensor core MMA reduction simultaneously for all the $m^2$ groups that can be made for the array $X[1..n]$. Lastly, once the reduction of a group is done, the thread in charge writes this result in its corresponding location in $X$, taking another unit of time. Combining these costs into one expression leads to the following recurrence 
\begin{equation}
    T_{tc}(n) = 5 + T_{tc}(\frac{n}{m^2})
\end{equation}
that stops at $T_{tc}(m^2) = 5$. Solving the recurrence leads to the final cost of
\begin{equation}
    T_{tc}(n) = 5\log_{m^2}(n)
\end{equation}
For comparison, a typical GPU reduction, under the same cost model, would take unit of time for reading the first element of each thread,  another unit of time for reading the second element, a unit of time for adding two numbers, and another unit of time for storing the value back to memory. The recurrence for such algorithm would be $T(n) = 4 + T(n/2)$ which solves into $T(n) = 4\log_2(n)$. Based on this cost model, the potential speedup for the tensor core based reduction would become
\begin{equation}
    S = \frac{T(n)}{T_{tc}(n)} = \frac{4\log_2(n)}{5\log_{m^2}(n)} = \frac{4}{5}\log_2(m^2).
\end{equation}
A value of $m \ge 2$, which is the minimum value $m$ could take in order for the reduction to work, is already sufficient to provide a result of $S > 1$.

\section{Discussion and Conclusions}
\label{sec_discussion}
The main contribution of this work is the presentation and analysis of a new tensor core based reduction based on matrix-multiply-accumulate operations. The intuition behind the algorithm lies in defining each iteration of the parallel reduction as a sequence of two MMA operations that act together to provide the summation for groups of $m^2$ values. In the first MMA operation the elements are inserted in the $A$ matrix, while $B$ is all-one and $C$ is a zero matrix. In the second MMA operation, $A$ is an all-one matrix, $B$ is the result of the previous MMA, and $C$ is zero again.

The main results obtained from the analysis are the time cost of the algorithm, which is $T_{tc}(n) = 5\log_{m^2}(n)$, and the speedup with respect to a classic parallel GPU reduction, in which we obtained a factor of $S = (4/5)\log_2(m^2)$. It is interesting to note that the minimum value that makes the reduction work, \textit{i.e.} $m=2$, already makes $S > 1$ and leads to a favorable result. Current GPUs perform $4\times 4 \times 4$ MMA operations in one cycle at hardware level, therefore there is an important chance that the potential speedup one could observe with current GPUs, such as Tesla V100 or Titan V, would be $S \approx 3.2$ if we consider $m=4$, and $S \approx 6.4$ if we consider $m=16$ which is the value exposed to the programmer. In either case, the performance speedup is significant and would contribute greatly to the problem of parallel reductions.

The results obtained in this work have shown that there is an important amount of potential performance that can be exploited from tensor cores, with applications in fields that are not necessarily related to machine learning. Indeed, it remains unknown what is the level of precision loss by performing reductions in FP16, and future work should include experimental results validating both the speedup values and the $\%$ of precision loss observed in different types of reductions.

\bibliographystyle{plain}
\bibliography{tensor-reduction}

\begin{thebibliography}{10}

\bibitem{case_study_omp_reduce}
Mahwish Arif and Hans Vandierendonck.
\newblock A case study of openmp applied to map/reduce-style computations.
\newblock In Christian Terboven, Bronis~R. de~Supinski, Pablo Reble, Barbara~M.
  Chapman, and Matthias~S. M{\"u}ller, editors, {\em OpenMP: Heterogenous
  Execution and Data Movements}, pages 162--174, Cham, 2015. Springer
  International Publishing.

\bibitem{Brent_1974}
Richard~P. Brent.
\newblock The parallel evaluation of general arithmetic expressions.
\newblock {\em J. ACM}, 21(2):201--206, April 1974.

\bibitem{CARTER2018148}
Francisco Carter, Nancy Hitschfeld, Crist{\'o}bal~A. Navarro, and Rodrigo Soto.
\newblock Gpu parallel simulation algorithm of brownian particles with excluded
  volume using delaunay triangulations.
\newblock {\em Computer Physics Communications}, 229:148 -- 161, 2018.

\bibitem{CERDA20188}
Mauricio Cerda, Crist{\'o}bal~A. Navarro, Juan Silva, Scott~R. Waitukaitis,
  Nicolás Mujica, and Nancy Hitschfeld.
\newblock A high-speed tracking algorithm for dense granular media.
\newblock {\em Computer Physics Communications}, 227:8 -- 16, 2018.

\bibitem{Chaitanya2017}
Chakravarty Reddy~Alla Chaitanya, Anton Kaplanyan, Christoph Schied, Marco
  Salvi, Aaron Lefohn, Derek Nowrouzezahrai, and Timo Aila.
\newblock Interactive reconstruction of noisy monte carlo image sequences using
  a recurrent autoencoder.
\newblock {\em ACM Trans. Graph.}, 36(4), 2017.

\bibitem{mapreduce}
Jeffrey Dean and Sanjay Ghemawat.
\newblock Mapreduce: Simplified data processing on large clusters.
\newblock In {\em Proceedings of the 6th Conference on Symposium on Operating
  Systems Design \& Implementation - Volume 6}, OSDI'04, pages 10--10,
  Berkeley, CA, USA, 2004. USENIX Association.

\bibitem{harris_2005}
Mark Harris.
\newblock Mapping computational concepts to gpus.
\newblock In {\em ACM SIGGRAPH 2005 Courses}, SIGGRAPH '05, New York, NY, USA,
  2005. ACM.

\bibitem{0031-9155-56-22-002}
Xun Jia, Xuejun Gu, Yan~Jiang Graves, Michael Folkerts, and Steve~B Jiang.
\newblock Gpu-based fast monte carlo simulation for radiotherapy dose
  calculation.
\newblock {\em Physics in Medicine \& Biology}, 56(22):7017, 2011.

\bibitem{tpu_google_2017}
Norman~P. Jouppi, Cliff Young, Nishant Patil, David Patterson, Gaurav Agrawal,
  Raminder Bajwa, Sarah Bates, and et~al.
\newblock In-datacenter performance analysis of a tensor processing unit.
\newblock {\em SIGARCH Comput. Archit. News}, 45(2):1--12, June 2017.

\bibitem{tensor_cores_2018}
Stefano Markidis, Steven Wei~Der Chien, Erwin Laure, Ivy~Bo Peng, and
  Jeffrey~S. Vetter.
\newblock {NVIDIA} tensor core programmability, performance {\&} precision.
\newblock {\em CoRR}, abs/1803.04014, 2018.

\bibitem{8392762}
C.~A. Navarro, M.~Vernier, N.~Hitschfeld, and B.~Bustos.
\newblock Competitiveness of a non-linear block-space gpu thread map for
  simplex domains.
\newblock {\em IEEE Transactions on Parallel and Distributed Systems}, pages
  1--1, 2018.

\bibitem{navarro_hitschfeld-kahler_mateu_2014}
Crist{\'o}bal~A. Navarro, Nancy Hitschfeld-Kahler, and Luis Mateu.
\newblock A survey on parallel computing and its applications in data-parallel
  problems using gpu architectures.
\newblock {\em Communications in Computational Physics}, 15(2):285–329, 2014.

\bibitem{NAVARRO201648}
Crist{\'o}bal~A. Navarro, Wei Huang, and Youjin Deng.
\newblock Adaptive multi-gpu exchange monte carlo for the 3d random field ising
  model.
\newblock {\em Computer Physics Communications}, 205:48 -- 60, 2016.

\bibitem{nickolls_2008}
John Nickolls, Ian Buck, Michael Garland, and Kevin Skadron.
\newblock Scalable parallel programming with cuda.
\newblock In {\em ACM SIGGRAPH 2008 Classes}, SIGGRAPH '08, pages 16:1--16:14,
  New York, NY, USA, 2008. ACM.

\bibitem{nvidia-cublas}
nVidia.
\newblock {\em {CUBLAS Library User Guide}}.
\newblock nVidia, v5.0 edition, October 2012.

\bibitem{cuda}
{NVIDIA Corporation}.
\newblock {NVIDIA CUDA C} programming guide, 2018.

\bibitem{omp_ompi}
R.~Rabenseifner, G.~Hager, and G.~Jost.
\newblock Hybrid mpi/openmp parallel programming on clusters of multi-core smp
  nodes.
\newblock In {\em 2009 17th Euromicro International Conference on Parallel,
  Distributed and Network-based Processing}, pages 427--436, Feb 2009.

\bibitem{SCHMIDHUBER201585}
J{\"u}rgen Schmidhuber.
\newblock Deep learning in neural networks: An overview.
\newblock {\em Neural Networks}, 61:85 -- 117, 2015.

\end{thebibliography}
\vspace{12pt}
\end{document}